\documentclass[aps,prstab,reprint,groupedaddress]{revtex4-1}

\usepackage{graphicx}% Include figure files
\usepackage{dcolumn}% Align table columns on decimal point
\usepackage{bm}% bold math
\usepackage{units}

\usepackage{todonotes}
\usepackage[caption=false]{subfig}

\usepackage{hyphenat}
\usepackage{color}

\begin{document}

\global\long\def\d{\mathrm{d}}
\global\long\def\en{\varepsilon_{0}}
\global\long\def\e{\mathrm{e}}
\global\long\def\ii{\mathbf{\imath}}

%\preprint{}


\title{FAST MAPPING OF TERAHERTZ BURSTING THRESHOLDS AND CHARACTERISTICS AT SYNCHROTRON LIGHT SOURCES}



\author{Miriam Brosi}\email{miriam.brosi@kit.edu}
\author{Johannes L. Steinmann}
\author{Edmund Blomley}
\author{Erik Br\"undermann}
\author{Nicole Hiller}\altaffiliation{now at PSI, Villigen, Switzerland}
\author{Benjamin Kehrer}
\author{Michael J.\ Nasse}
\author{Manuel Schedler}
\author{Patrik Sch\"onfeldt}
\author{Marcel Schuh}
\author{Markus Schwarz}
\author{Anke-Susanne M\"uller}
\affiliation{Karlsruhe Institute of Technology, LAS and IBPT}
\author{Michele Caselle}
\author{Lorenzo Rota}
\author{Marc Weber}
\affiliation{Karlsruhe Institute of Technology, IPE}


\date{\today}

\begin{abstract}

Dedicated optics with extremely short electron bunches enable synchrotron
light sources to generate intense coherent THz radiation. The high degree 
of spatial compression in this so-called low-$\alpha_{c}$ optics entails
a complex longitudinal dynamics of the electron bunches, which can be 
probed studying the fluctuations in the emitted terahertz radiation 
caused by the micro-bunching instability (``bursting''). 
This article presents a ``quasi-instantaneous'' method for measuring the 
bursting characteristics by simultaneously collecting and evaluating the 
information from all bunches in a multi-bunch fill, reducing the 
measurement time from hours to seconds.
This speed-up allows systematic 
studies of the bursting characteristics for various accelerator settings 
within a single fill of the machine, enabling a comprehensive comparison 
of the measured bursting thresholds with theoretical predictions by 
the bunched-beam theory. 
This paper introduces the method and presents first results obtained at the ANKA synchrotron radiation facility. 

\end{abstract}

\pacs{no PACS numbers yet}
%\keywords{}

\maketitle
%\tableofcontents
\section{\label{sec:level1}Introduction}

Short intense pulses of coherent synchrotron radiation (CSR) in the
terahertz (THz) frequency range are generated at synchrotron radiation facilities 
when electron bunches are compressed to picosecond timescales.
In this operation mode the high degree of spatial compression of the optics with reduced momentum compaction factor (``low-$\alpha_{\mathrm{c}}$ optics'') 
entails complex longitudinal dynamics of the electron bunches, 
leading to the so-called micro-bunching instability. 
This causes 
time-dependent fluctuations and strong bursts in the radiated THz 
intensity and is characterized by a threshold current
per bunch, the so-called bursting threshold.

The bursting threshold in the generation of CSR was studied
at several synchrotron light sources, e.g. ANKA \cite{anke2010}, BESSY II \cite{bakr2003},   
DIAMOND \cite{diamond_bursting_2012}, MLS \cite{mls_ipac10}, 
NSLS VUV Ring \cite{carr1999} and SOLEIL \cite{soleil_2012}.
One of the first theoretical explanations of this threshold was  
given by Stupakov \cite{Stupakov2002}. Later work extended the 
theory by considering self-interactions due to CSR and shielding effects
due to the vacuum chamber, considering coasting or bunched electron beams
\cite{venturini2002, Stupakov2002, Bane_2010, cai2011ipac}.
First comparisons between the 
measured bursting threshold and theoretical models were done for 
the bunched-beam theory in \cite{cai2011ipac} and \cite{ries_ipac12}.

This article presents a novel ``quasi-instantaneous'' approach to study the 
bursting threshold, based on high-rate sampling over many turns of the 
radiation emitted from all electron bunches circulating in the ANKA storage 
ring. These so-called ``snapshot measurements'' drastically reduce the time 
necessary for mapping the bursting behavior and the bursting threshold at 
different machine settings. The measured bursting thresholds at these different 
configurations are compared to the bunched-beam theory~\cite{Bane_2010}, which is briefly 
outlined below.

\section{Bunched-Beam theory}\label{sec:th}

For wavelengths longer than the emitting charge structure, synchrotron
radiation is emitted coherently with a spectral power proportional to 
the number of electrons squared, the power spectrum of a single electron 
and a form factor, which is the modulus squared of the Fourier transform 
of the normalized charge distribution of the radiating electron 
bunch \cite{anke2012}. If sufficiently narrow sub-structures exits, 
CSR can be emitted at shorter wavelengths than expected from the natural 
bunch length. In a storage ring, above a certain threshold in 
the bunch current, the CSR impedance causes a modulation of the longitudinal
phase space, which gives rise to time-evolving sub-structures in the 
longitudinal particle distribution. As a consequence of continuous changes 
in the form factor, this micro-bunching instability \cite{Stupakov2002} 
leads to strong fluctuations in the emitted CSR power, referred to as 
bursting, which occur with characteristic frequencies. Formally
the occurrence of sub-structures in the longitudinal phase space above the
bursting threshold can be described by the Vlasov-Fokker-Planck equation
\cite{venturini2002}, taking into account the geometry of the storage 
ring and accelerator parameters, like the accelerating voltage and the 
momentum compaction factor.

The bunched-beam theory \cite{Bane_2010} describes the influence of the 
vacuum chamber on a bunched electron beam. The resulting condition for the
bursting threshold, $\left(S_{\mathrm{CSR}}\right)_{\mathrm{th}}$, as a function of the shielding parameter $\Pi$, 
in the following taken from the parallel plates model \cite{Bane_2010}
for a vacuum chamber of height $2 h$, is found to be
\begin{equation}
\left(S_{\mathrm{CSR}}\right)_{\mathrm{th}}\left(\Pi\right)=0.5+0.12\,\Pi
\quad\mbox{with}\quad
\Pi=\frac{\sigma_{\mathrm{z},0}R^{\frac{1}{2}}}{h^{\frac{3}{2}}} \;.
\end{equation}
Here the threshold is expressed in terms of the CSR strength 
$S_{\mathrm{CSR}}=I_{\mathrm{n}}R^{1/3}/\sigma_{\mathrm{z},0}^{4/3}$, with
$I_{\mathrm{n}}=\sigma_{\mathrm{z},0}I_{\mathrm{b}}/(\alpha_{c}\gamma\sigma_{\delta}^{2}I_{\mathrm{A}})$, 
the normalized bunch current, $I_{\mathrm{b}}$ the bunch current, $R$ 
the bending radius, $\sigma_{\mathrm{z},0}$ the natural bunch length, $\alpha_{\mathrm{c}}$ 
the momentum compaction factor, $\sigma_{\delta}$ the relative energy 
spread, $\gamma$ the Lorentz factor and 
$I_{\mathrm{A}}=4\pi\varepsilon_{0}m_{\mathrm{e}}c^{3}/e=\unit[17045]{A}$ the Alfv\'en current.
The momentum compaction factor and the natural bunch length are 
determined by the beam energy $E$, the synchrotron frequency $f_\mathrm{s}$, 
the RF frequency $f_{\mathrm{RF}}$, the revolution frequency $f_{\mathrm{rev}}$, the RF peak 
voltage $V_{\mathrm{RF}}$ and the radiated energy per particle and revolution
$U_0$ as
\begin{equation}
\alpha_{\mathrm{c}}
=\frac{E f_{\mathrm{s}}^{2} 2\pi}{f_{\mathrm{RF}} f_{\mathrm{rev}} \sqrt{e^{2}V_{\mathrm{RF}}^{2} - U_{0}^{2}}}
 \quad\mbox{and}\quad
 \sigma_{\mathrm{z},0}= \frac{\alpha_{\mathrm{c}} \sigma_\delta}{2\pi f_{\mathrm{s}}} \;.
 \label{eq:momcomp}\label{eq:bunchlength}
\end{equation}

The above relations allow to express the bunch current at the bursting 
threshold 
\begin{equation}
I_{\mathrm{b}}^{\mathrm{th}} = I_{\mathrm{A}} \gamma \sigma_{\delta}^{2} \alpha_{\mathrm{c}} 
        R^{-\frac{1}{3}} \sigma_{\mathrm{z},0}^{\frac{1}{3}}
       \left( 0.5 + 0.12 R^{\frac{1}{2}} \sigma_{\mathrm{z},0} h^{-\frac{3}{2}}\right) 
\label{eq:burstingschwelle}
\end{equation} 
as a function of parameters characterizing the longitudinal beam dynamics: 
the momentum compaction factor, the energy spread and the natural bunch 
length and, therefore, the RF-voltage and the synchrotron frequency. Likewise,
using these dependencies, studies of the bursting threshold allow
non-invasive  diagnostics of the beam dynamics in longitudinal phase space.

\section{Experimental setup}

The core of the ANKA synchrotron radiation source of the Karlsruhe Institute of Technology, Germany, is a 110.4 m long electron storage ring, operating in the energy range from 0.5 GeV to 2.5 GeV. 
The low-$\alpha_{\mathrm{c}}$ mode at ANKA allows the reduction of the bunch length down to a few picoseconds \cite{anke2005}. A state-of-the-art bunch-by-bunch feedback system \cite{hertle_ipac14} is used to generate custom filling patterns (e.g., a single bunch, arbitrary bunch distances and varying bunch currents) required for the beam dynamics studies presented in this paper. 

For the investigations of the micro-bunching instability at ANKA, a variety of fast THz detector systems are used in combination with fast, in-house built, high-repetition-rate data acquisition (DAQ) systems. 
For the bunch-by-bunch, turn-by-turn studies of the rapidly fluctuating THz radiation, fast detectors with different characteristics, e.g., sensitivity, or noise figures, are available at KIT. 
One example is an in-house developed THz detector system based on a YBCO sensor~\cite{Raasch_2015}. 
The investigations of this paper were performed with a broad band quasi-optical Schottky diode (ACST GmbH, Hanau, Gemany) with a sensitivity in the spectral range from several \unit[10]{GHz} up to \unit[2]{THz} with the peak sensitivity around \unit[80]{GHz} \cite{acst_flyer}. 
With a bunch spacing at ANKA of \unit[2]{ns}, corresponding to the maximum bunch repetition rate of \unit[500]{MHz}, the built-in \unit[4]{GHz} amplifier of the detector is fast enough to resolve the THz pulse of each bunch individually.

The KArlsruhe Pulse Taking and Ultrafast Readout Electronics system (KAPTURE) \cite{caselle_ipac14}, developed at KIT, implements a memory-efficient approach to acquire the detector signal on a bunch-by-bunch basis. 
The signal of the fast THz detector is fed via high bandwidth connectors into a track-and-hold unit and a 12-bit, \unit[500]{MSa/s} % TODO: MS/s?
 analog-to-digital converter. KAPTURE offers up to 4 sampling channels with individual delay units, whose sampling trigger points can be adjusted in \unit[3]{ps} steps each.
This enables a ``local sampling'' of a detector signal in minimum steps of \unit[3]{ps}. 
This local sampling depends on the chosen delays between the channels, with a maximum rate of \unit[330]{GSa/s}, compared to a global sampling rate of approximately \unit[500]{MSa/s} in each channel \cite{caselle_joi2014}.

For the studies detailed below, the fast THz detector measures the THz pulse emitted by each bunch at each turn. 
The DAQ system, KAPTURE,  was configured to record only the amplitude of the detector response to the THz pulse.  
During a measurement period of one second, the amplitude of the detector pulses for each individual bunch is thus recorded for approximately 2.7 million consecutive turns. In the following, we will refer to this data as the THz signal of each bunch.

\section{Beam studies}

The threshold current for bursting emission of CSR depends on several (key) beam parameters, such as the momentum compaction factor and the natural bunch length (see Eq.~\ref{eq:momcomp}). It can therefore be used to access the longitudinal beam dynamics. 
Furthermore, knowledge of the bunch currents where the micro-bunching instability occurs for a given set of machine parameters is also critical for experiments that rely on stable THz emission.

\subsection{Bursting THz radiation}
\label{sec:slow_technique}

\begin{figure}[!htb]
 \includegraphics[width=0.5\textwidth]{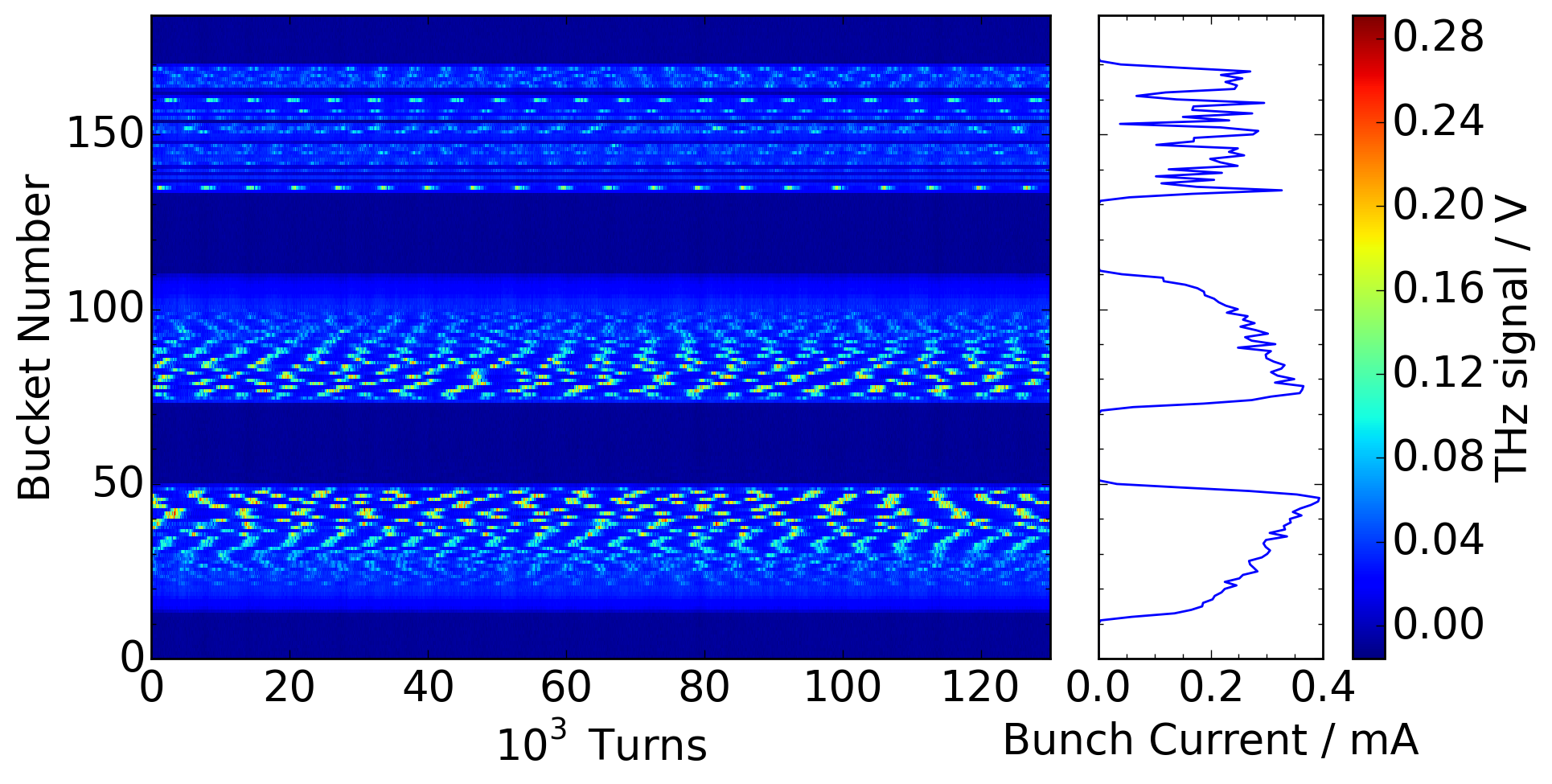}
% {f05110_2014-04-16_SD_squeeze-scan_15_float32_file002_1397680452_plot_ADC_1_notitle.png}%
 \caption{The fluctuating THz signal (color coded) of each of the 184 RF-buckets is shown for the first 130 thousand consecutive turns, out of the total recorded of 2.7 million in one second. 
The bursting behavior differs for bunches with different currents.
On the right hand side the filling pattern, consisting of three trains, is indicated by the bunch current (adapted from \cite{brosi_ipac15}).}
\label{fig:dataset}
\end{figure}

With the setup described above, the amplitude of the detector response for the
individual THz pulse of each bunch is recorded for every turn of over 2.7
million consecutive revolutions. Such a dataset contains the THz trace 
of all bunches during the measurement time of one second. The first 130 
thousand turns of one such dataset are displayed in Fig.~\ref{fig:dataset}. 
Each horizontal row shows the THz signal as a function of the turn number. 
A vertical column yields the THz signal for all bunches at a specific turn. 
The filling pattern for this measurement consisted of three trains with
about 30 bunches each (Fig.~\ref{fig:dataset}). The bunch currents were 
measured with a time-correlated single-photon counting setup \cite{kehrer_ipac15}. 

The fluctuations in the intensity due to the bursting are clearly visible 
for each bunch. Bunches with different currents display a different
bursting behavior, i.e., temporal evolution of a burst and its 
repetition rate. To measure these current-dependent changes, a dataset 
was taken every 10 seconds while the total beam current decreased. 
Combined with the measured bunch currents, these measurements allow to map fluctuations 
of the intensity emitted by each bunch as a function of the turn number
($T_{\mathrm{rev}}=\unit[368]{ns}$ for ANKA) and of the decreasing bunch current 
(time scale typically between one and several hours).

\begin{figure}[!htb]
 \includegraphics[width=0.5\textwidth]{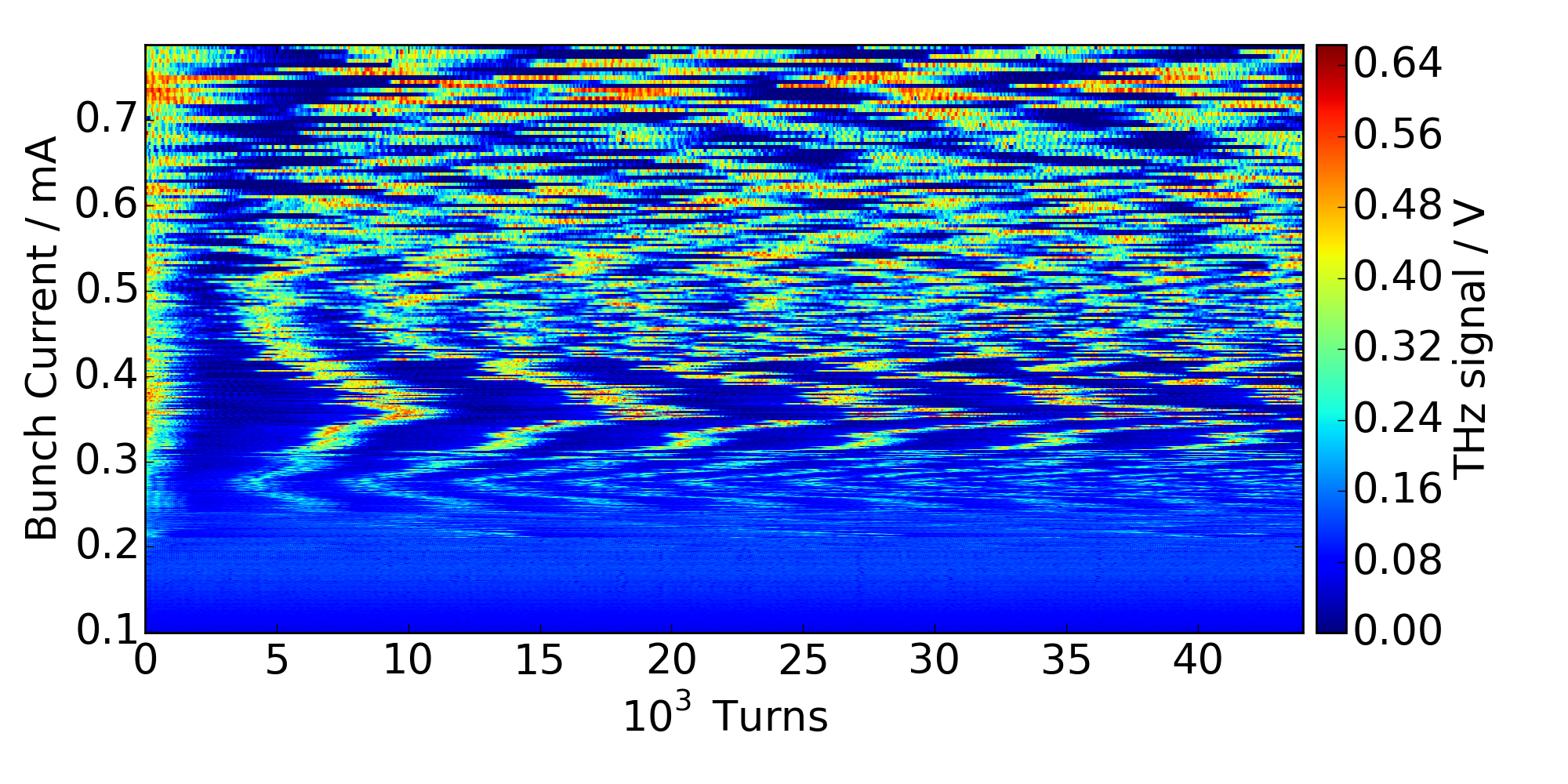}
% {f04922_2013-11-27_SD_2schottky_decay_ADC0_float32_int_waterfall_bucket155_601files_ADC0_mavr_linear.png}%
 \caption{THz signal (color coded) of one bunch as a function of turn number during the decrease of the bunch current. Both the temporal evolution and the repetition rate of an outburst show a strong dependence on bunch current. For better visibility the THz signals were shifted in time so that the first bursts are aligned.}
\label{fig:int_waterfall}
\end{figure}

In Fig.~\ref{fig:int_waterfall} the THz signal of one individual bunch is displayed as a function of the decreasing bunch current, showing changes in the bursting behavior as well as the onset of bursting in the detection range of the detector system used for this study, in this case, at \unit[0.2]{mA}.

%%%\bibliography{references_thesis_bibdesk_loaded}
%%%\end{document}

\subsection{Fast mapping technique}

The measurement technique described in the previous section collects the THz signal traces of all 184 RF-buckets for one second, during which time the bunch currents stay approximately constant. 
This way, a study of current dependent effects is usually achieved by repeating those one-second-measurements of THz signal traces for changing beam currents. To avoid possible influences from continued beam injection, or in the absence of a full energy injection, the measurement series are performed while the beam current slowly decays due to a limited beam lifetime over a time span of several hours.   

The fast mapping technique (the so-called "snapshot" measurement), enabled by the unique combination of fast THz detectors with high-data-throughput DAQ systems, 
drastically reduces the time required to cover the full bunch current range of interest: instead of following one single bunch during an hour-long beam current decrease, the technique makes use of the quasi-simultaneous acquisition of all bunches, for this purpose filled to cover the full current range of interest, to achieve the same result in one second.  
The special, tailored filling pattern required is achieved with the help of a bunch-by-bunch feedback system \cite{hertle_ipac14}.

With such a filling pattern, a snapshot measurement of one dataset within one second is sufficient to analyze the bursting behavior at different bunch currents and to determine the bursting threshold corresponding to the present machine settings. 
Measurements, which previously lasted typically up to 5 hours, depending on the lifetime of the electron beam, now only require one second or less. 

\subsubsection{Mapping the bursting behavior}

\begin{figure*}[!tb]
    \centering
   \subfloat[]{
        \includegraphics[width=0.5\textwidth]{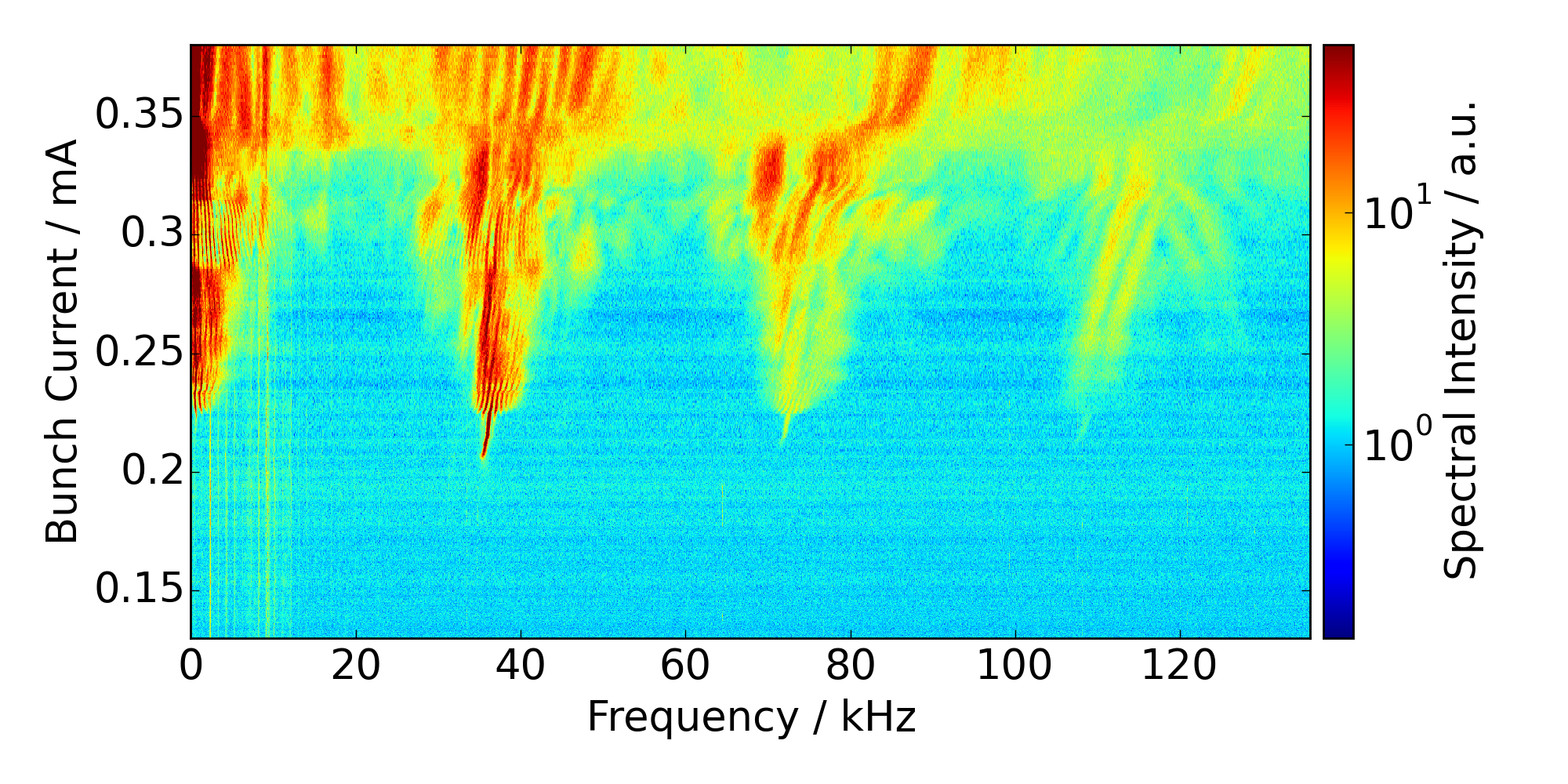}
%        {f05135_2014-05-07_SD_overnight_ADC2_float32_spectrogram_bucket130_551files_ADC2_mavr10_0-150khz_013-038mA.png}
        \label{fig:spectrogram_2}}%
     ~     
         \subfloat[]{
        \includegraphics[width=0.5\textwidth]{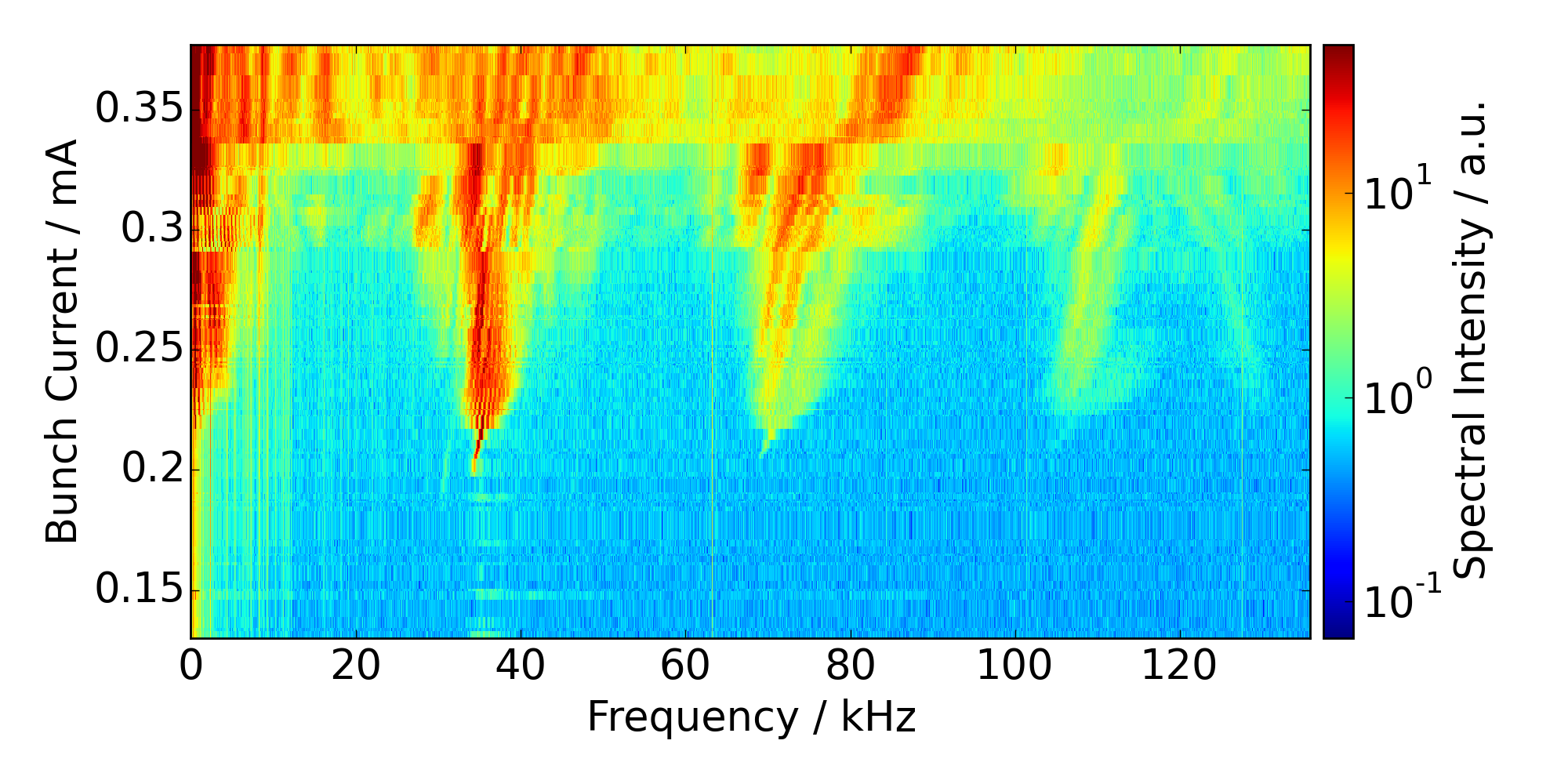}
%        {f05110_2014-04-16_SD_squeeze-scan_15_float32_file002_1397680452_instant_waterfall_ADC_1_mavr10_0-150khz_notitle_linearcurrent_125-100000.png}
        \label{fig:instant_spectrogram}}
\caption{The spectrogram shown in (a) was obtained from measurements of a single bunch for slowly changing beam current lasting two and a half hours, while the spectrogram in (b) was obtained from a snapshot measurement lasting just one second by using the FFT of THz signals similar to those shown in Fig.~\ref{fig:dataset}. For the snapshot measurement the bunch current resolution is limited by the number of bunches and their current distribution. Compared to this, the standard measurement during a slow current decrease results in a higher current resolution. Despite the limited resolution of the snapshot spectrogram, the dominant bursting frequencies and the thresholds between different bursting regimes are clearly visible.} 
     \label{fig:spectrogram_comparison}
\end{figure*}

The snapshot measurements offer the opportunity to acquire fast, comprehensive maps of the micro-bunching instability in dependence of various accelerator parameters. 
To visualize the change in bursting periodicity as a function of the bunch current, the Fourier transform of the THz signal is displayed for the investigated bunch current range in a spectrogram (see Fig.\ref{fig:spectrogram_comparison}).
Bunch currents that mark a change in the fluctuation of the THz signal and therefore in the driving micro-bunching instability are easily accessible in the visualization  \cite{schwarz2013ipac}. 
With the quasi-simultaneous acquisition of bunches covering the full bunch current range, a spectrogram can now be created from just one dataset.

The resulting spectrogram has a limited resolution on the current axis, due to the limited number of bunches and hence current bins, as shown in Fig.~\ref{fig:instant_spectrogram} in comparison to the spectrogram obtained from measurements of a single bunch for different bunch currents shown in Fig.~\ref{fig:spectrogram_2}. 
However, the different dominant frequencies and regions are clearly visible and give an overview of the bursting behavior for these accelerator settings. 
A comparison between the snapshot spectrogram and a spectrogram taken during a standard current decay of one bunch at the same accelerator settings (Fig.~\ref{fig:spectrogram_comparison}) shows the excellent agreement of the two methods and that the dominant structures can easily be reproduced by the time-saving snapshot measurement.

\subsubsection{Fast threshold determination}

\begin{figure}[!b]
 \includegraphics[width=0.5\textwidth]{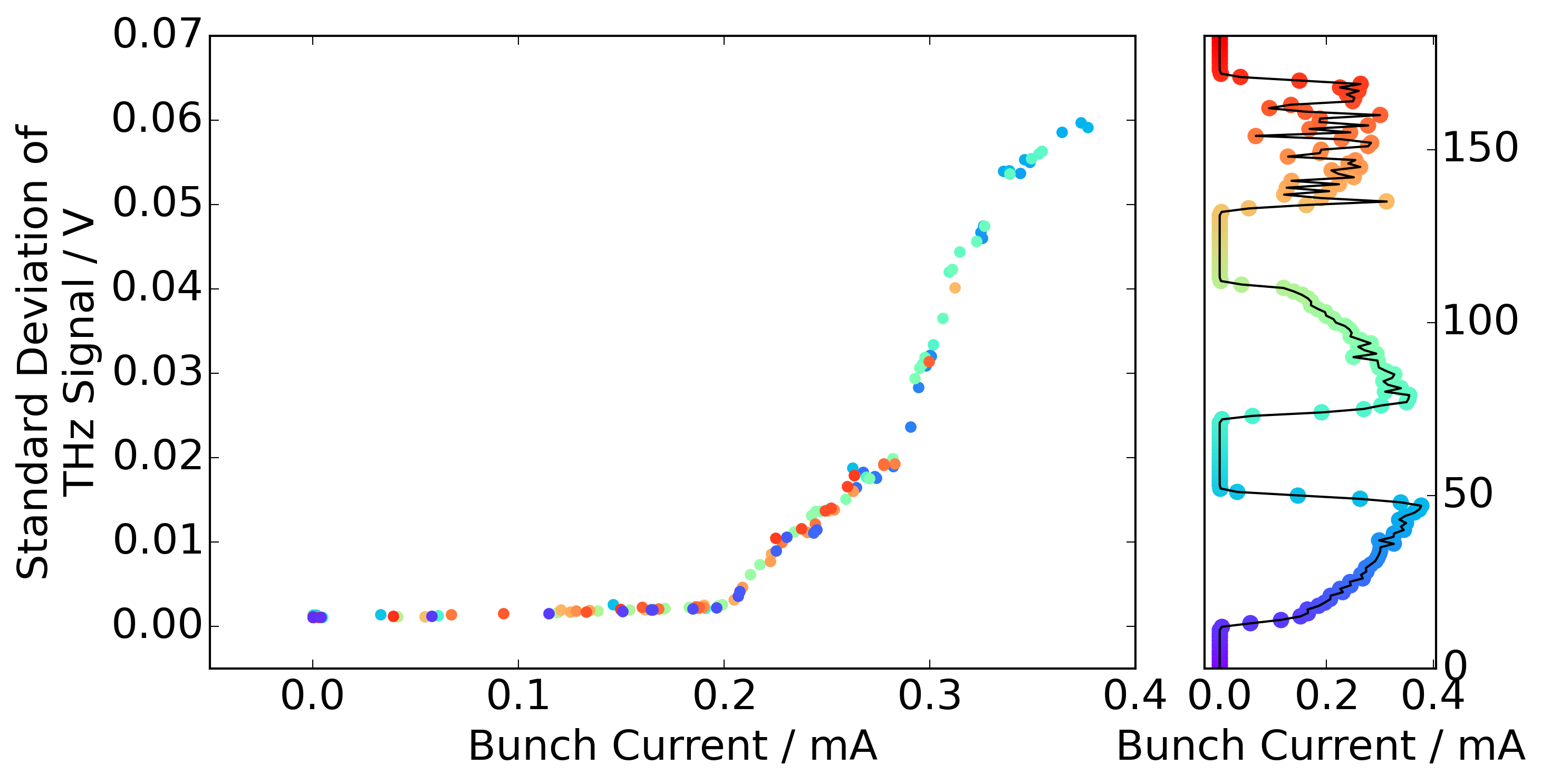}
% {f05110_2014-04-16_SD_squeeze-scan_15_float32_timedomain_intensity_over_current_bucket0-183_file2_ADC_1_1397680452_deviation_linear_linepoints.png}%
% 
 \caption{Standard deviation of the THz signal of each bunch as a function of the corresponding bunch current, revealing the onset of bursting emission (``bursting threshold'') at \unit[0.2]{mA}. The individual bunch currents (shown on the right) were patterned to equally sample the whole bunch current range (adapted from \cite{brosi_ipac15}).}
 \label{fig:squeeze-scan_std}
\end{figure} 

To extract the bursting threshold from a snapshot measurement, the standard deviation of the THz signal of each bunch is calculated and displayed as a function of the momentary current of the bunch in question (Fig.~\ref{fig:squeeze-scan_std}).
The bursting threshold is immediately visible as a kink in the standard deviation, because the THz signal is stable at currents below the threshold. Above the instability threshold, when the THz signal starts to fluctuate over the turns, the value of the standard deviation increases. 

 Figure~\ref{fig:squeeze-scan_std} shows that the THz signals of bunches below the threshold of \unit[0.2]{mA} show hardly any fluctuations compared to the signal of bunches with higher current.
 \linebreak

\subsubsection{Bursting threshold and beam optics}
\label{par:squeeze-scan_over_fs}

The bursting threshold $I^{\mathrm{th}}_{\mathrm{b}}$ depends on the momentum compaction factor
$\alpha_{\mathrm{c}}$, both directly (Eq.~\ref{eq:burstingschwelle}) and indirectly
through the natural bunch length $\sigma_{\mathrm{z},0}$. Through $\alpha_{\mathrm{c}}$ and $\sigma_{\mathrm{z},0}$ it also 
depends indirectly on the RF-voltage $V_{\mathrm{RF}}$ and the synchrotron
frequency $f_{\mathrm{s}}$ (Eq.~\ref{eq:momcomp}).

\begin{figure}[!t]
 \includegraphics[width=0.472\textwidth]{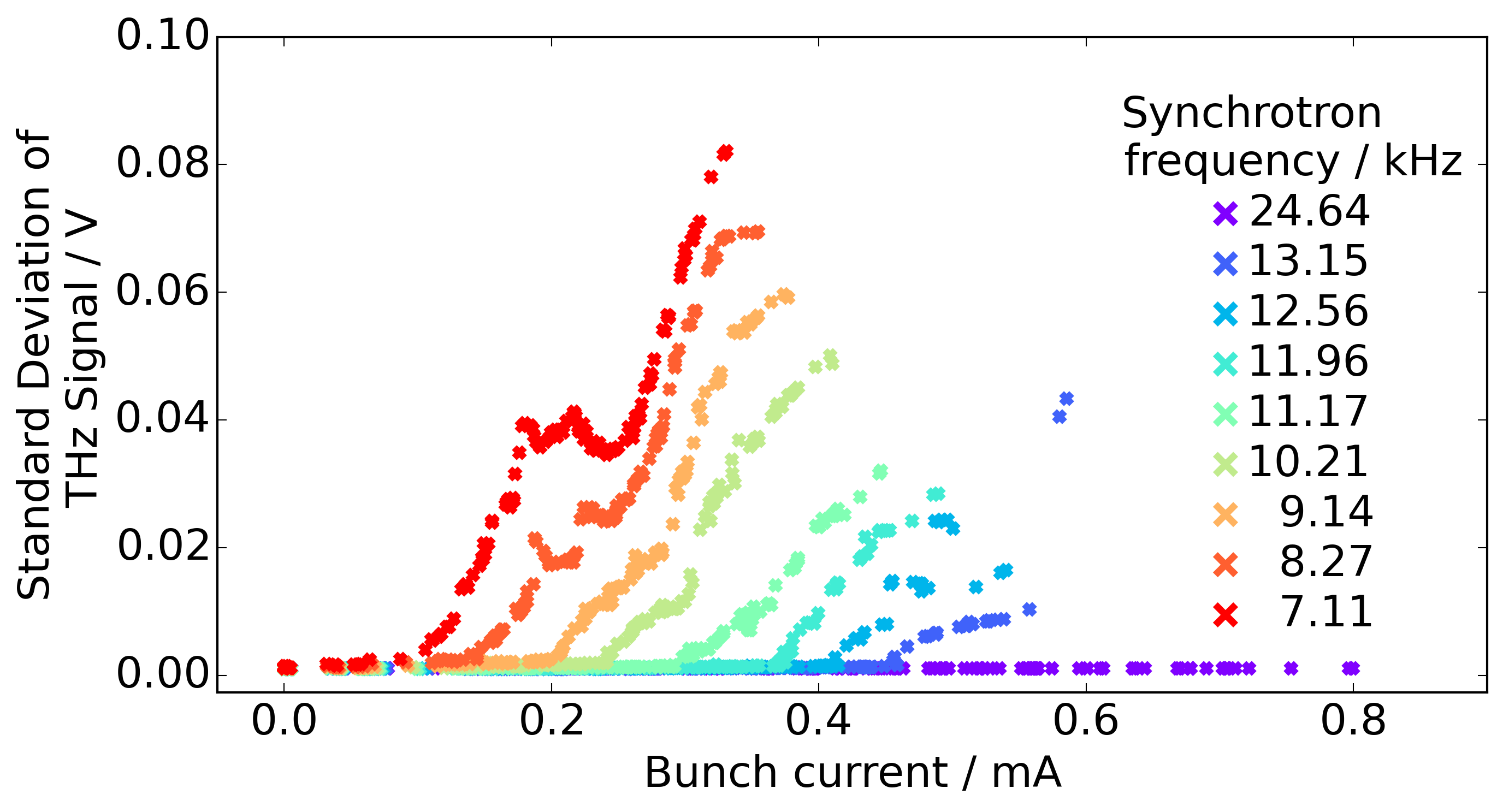}
% {f05110_2014_04_16_SD_squeeze-scan-deviation-Intensity-current_files_19-29ksteps_points_wo-khz.png}%
 \caption{Standard deviation of the THz signal as a function of the bunch
   current for a constant RF-voltage of \unit[1047]{kV} and different 
   settings of the magnet optics, resulting in different synchrotron frequencies
   and hence different bursting thresholds (adapted from \cite{judin_ipac14}).
 \label{fig:squeeze-scan_std-all}}
\end{figure}

To study these dependencies, snapshot measurements were used to 
determine the bursting threshold for different settings of the magnet 
optics, resulting in different values of the momentum compaction 
factor at a fixed RF-voltage. In the following, the stepwise
change of the magnet optics will be referred to as ``magnet sweep''. 
At each step a snapshot measurement was taken. 
Figure~\ref{fig:squeeze-scan_std-all} shows the standard deviation of
the intensity of the THz signal as a function of the bunch current for several 
steps of the sweep. With decreasing synchrotron frequency $f_{\mathrm{s}}$ a change 
in the threshold, i.e., the onset of the fluctuations, is visible. 

\begin{figure*}[!htb]
\includegraphics[width=1\textwidth]{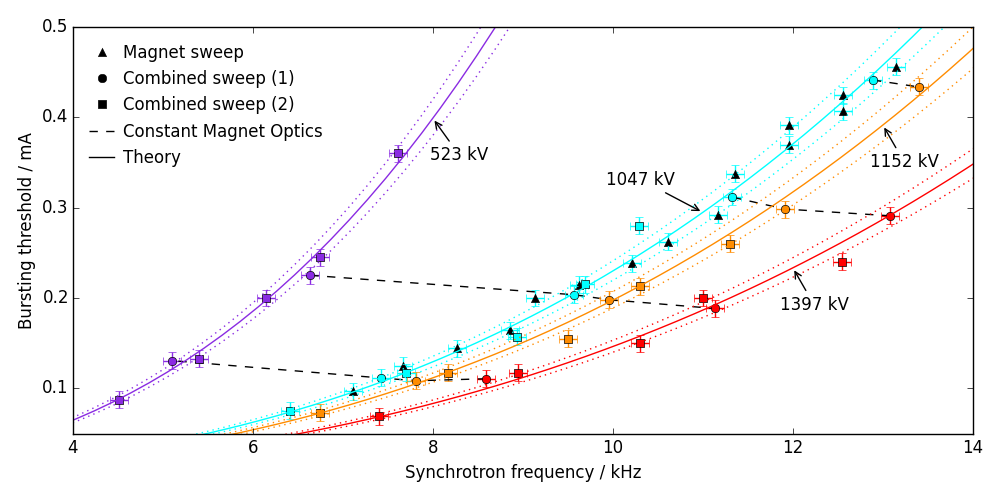}
%{i_th_over_fs_both_black-triangle-and-lines.png}%
 \caption{Measured bursting thresholds from three sets of measurements
   as a function of the synchrotron frequency. Here ``combined sweep (2)'' 
   refers to the same measurements taken at 11 months apart from ``combined sweep (1)''. For
   ``combined sweep (1)'', measurement with the same magnet optics are connected
   by dashed lines. The different colors indicate the different acceleration
   voltages. The horizontal error bars indicate the systematic error on the
   measured synchrotron frequency. The vertical error bars include the error of
   the algorithm for the automated threshold detection, the bunch current
   measurement as well as the expected spread of the threshold due to
   multi-bunch effects. The solid lines show the theoretical expectation
   according to Eq.~\ref{eq:burstingschwelle} from the bunched-beam theory. The
   dotted lines indicate the one standard deviation uncertainties on the 
   theoretical predictions. Within uncertainties all measurements agree 
   with the bunched-beam theory. \label{fig:i_th_over_fs_with_theory}}
\end{figure*}

Displaying the bursting threshold in dependence of the measured synchrotron frequency
$f_{\mathrm{s}}$ for each step of the magnet sweep shows the correlation even more clearly
(Fig.~\ref{fig:i_th_over_fs_with_theory}). The decrease of the threshold with 
decreasing synchrotron frequency matches the expected  behavior. 
For a fixed acceleration voltage the synchrotron frequency decreases for a 
lower momentum compaction factor, which leads to a 
shorter bunch length. For a shorter bunch 
length the critical charge density above which the micro-bunching instability 
occurs is reached at a lower current. For a fixed RF-voltage, a lower 
synchrotron frequency corresponds to a lower bursting threshold 
(Fig.~\ref{fig:i_th_over_fs_with_theory}).
 
The error on the threshold contains the uncertainties of the  threshold
detection algorithm, the measured bunch current and the error expected by the spread of
the threshold due to multi-bunch effects \cite{brosi_ipac15}. \hfil

% \linebreak

\subsubsection{Bursting threshold, beam optics, and RF-voltage}
\label{par:squeeze-rf-scan_over_fs}

In a second measurement the RF-voltage was changed stepwise together  
with the magnet optics, in the following referred to as ``combined sweep''. 
Figure~\ref{fig:i_th_over_fs_with_theory} shows the measured bursting
thresholds as a function of the measured synchrotron frequency. For an RF-voltage 
of $\unit[1047]{kV}$ the thresholds are in good agreement with the 
thresholds seen in the magnet sweep study 
(Sec.~\ref{par:squeeze-scan_over_fs}), where the same voltage was used.

As described above, a bunch length reduction due to a change of 
the magnet optics that leads to a lower momentum compaction factor, 
decreases the threshold. 
Furthermore, for a fixed momentum compaction factor, a shortening of the 
bunch by an increase in the RF-voltage which is accompanied by an  
increase of the synchrotron frequency (Eq.~\ref{eq:momcomp}) also 
results in a decrease of the threshold. For the second effect the relative
changes of the threshold with the synchrotron frequency are smaller 
than for the first one (Fig.~\ref{fig:i_th_over_fs_with_theory}).

\section{Comparison with bunched-beam theory}

In the following, the measured bursting thresholds are compared to absolute predictions of the bunched-beam theory. 
For this comparison the synchrotron frequency $f_{\mathrm{s}}$ as well 
as the accelerating voltage were determined for each setting during the 
measurements. The compact Compton backscattering setup at ANKA 
\cite{chang_ipac15} was used to determine the momentum compaction factor 
for several different magnet optics \cite{chang_diss}. This allowed the 
extraction of a voltage calibration factor, which relates the acceleration 
voltage $V_{\mathrm{RF}}$ seen by the electrons to the set values of the cavities. 
Knowledge of these parameters is then used to calculate the natural bunch
length for each sweep step, and from that the theoretically predicted bursting 
threshold according to Eq.~\ref{eq:burstingschwelle}.

Figure~\ref{fig:i_th_over_fs_with_theory} shows the measured bursting thresholds
of three different sets of measurements, the magnet sweep and two combined sweeps
(taken 11 months apart) as a function of the synchrotron frequency.
The theoretical prediction according to Eq.~\ref{eq:burstingschwelle} for the
four different acceleration voltages is shown, and the one standard deviation uncertainty on the
theoretical values, calculated from the errors on the measured input parameters used
in Eqs.~\ref{eq:bunchlength} and~\ref{eq:burstingschwelle}, is
indicated. Within the uncertainties, the measured thresholds agree extremely well with 
the theoretical values, derived from the model as detailed in Sec.~\ref{sec:th}. 

\section{Conclusion} 

Fast snapshot measurements, enabled by the unique combination of fast THz detectors with the high-data-throughput KAPTURE DAQ system, drastically speed up time consuming surveys of the micro-bunching instability for different accelerator settings. 
The quasi-simultaneous acquisition of THz signals from all bunches allows the determination of the bursting threshold and the calculation of a spectrogram. This spectrogram shows the characteristic 
bursting frequencies from a single dataset covering one second of beam time, 
compared to the previously used methods requiring several hours. This speed-up by about
4 orders of magnitude permits systematic studies of the bursting threshold for 
a large set of machine parameters within the same fill. The reproducibility of 
this method is confirmed by the consistency of the results from two similar 
measurements performed 11 months apart. With measurements of only three different 
fills, it was possible to gather sufficient data to demonstrate that the 
absolute predictions of the bunched-beam theory for the bursting threshold are in excellent agreement with the measurements. 

%\appendix
%\section{}

\begin{acknowledgments}

The authors would like to thank the infrared group at ANKA and in particular M. S\"upfle and Y.-L. Mathis for their support during the beam times at the IR1 and IR2 beam lines as well as Cheng Chang for the Compton backscattering measurements (CBS) and Heinz-Wilhelm H\"ubers and Heiko Richter (DLR, Berlin) for providing the necessary CO2 laser for the CBS measurements. Further, the authors would like to thank the ANKA THz group for inspiring discussions. This work has been supported by the German Federal Ministry of Education and Research (Grant No. 05K13VKA), and the Helmholtz Association (Contract No. VH-NG-320). 
Miriam Brosi acknowledges the financial support of the Helmholtz International Research School for Teratronics (HIRST).
\end{acknowledgments}

\appendix*

%\section{Appendixes}
%
%To start the appendixes, use the \verb+\appendix+ command.

%\nocite{*}

%\bibliography{references_thesis_bibdesk_loaded,dipl_patrik,diss_vitali}
\bibliography{references_thesis_bibdesk_loaded}

\end{document}